\documentclass[aps,prl,twocolumn,superscriptaddress,10pt]{revtex4-1}

\usepackage{graphicx}

\usepackage{amsmath}

\usepackage{color}

\usepackage{epstopdf}

\usepackage{time,mathptmx}

\begin{document}

\title{Observation of Self-Similarity in the Magnetic Fields Generated by the Ablative Nonlinear Rayleigh-Taylor Instability}

\author{L.~Gao$^{1,2}$, P.~M.~Nilson$^{1,3}$, I.~V.~Igumenschev$^{1}$,
G.~Fiksel$^{1}$, R.~Yan$^{1,2,3}$, J.~R.~Davies$^{1,2,3}$, D.~Martinez$^{4}$, V.~Smalyuk$^{4}$, M.~G.~Haines$^{5}$\footnote{Deceased}, 
\\E. G. Blackman$^{1,6}$, D.~H.~Froula$^{1}$, R.~Betti$^{1,2,3,6}$, and D.~D.~Meyerhofer$^{1,2,3,6}$}

\affiliation{$^{1}$Laboratory for Laser Energetics, University of Rochester, Rochester, NY, 14623, USA} 
\affiliation{$^{2}$Department of Mechanical Engineering, University of Rochester, Rochester, NY, 14623, USA}
\affiliation{$^{3}$Fusion Science Center for Extreme States of Matter, University of Rochester, Rochester, NY, 14623, USA}
\affiliation{$^{4}$Lawrence Livermore National Laboratory, Livermore, California 94550, USA}
\affiliation{$^{5}$Department of Physics, Imperial College, London SW7 2AZ United Kingdom}
\affiliation{$^{6}$Department of Physics and Astronomy, University of Rochester, 
Rochester, NY, 14623, USA}

\date{\today}


\begin{abstract}

\noindent Magnetic fields generated by the nonlinear Rayleigh-Taylor growth of laser-seeded three-dimensional broadband perturbations were measured in laser-accelerated planar targets using ultrafast proton radiography. The experimental data show self-similar behavior in the growing cellular magnetic field structures. These observations are consistent with a bubble competition and merger model that predicts the time evolution of the number and size of the bubbles, linking the cellular magnetic field structures with the Rayleigh-Taylor bubble and spike growth. 




\end{abstract}




\maketitle


Understanding the nonlinear behavior of matter under conditions of extreme temperature, pressure, and density is important for interpreting a wide range of high energy density phenomena \cite{NASHEDP,DrakeHEDP}. In the high energy density regime, a decisive role is played in many situations, from supernova hydrodynamics \cite{remington:1994} to inertial confinement fusion \cite{Atzeni2004,lindl1995}, by the formation of unstable flows \cite{Dimotakis}, the transition to turbulence \cite{Orszag1972}, and the creation of mixing layers \cite{Schneider1998}. Techniques are widely sought to access these conditions and investigate the high-temperature magnetohydrodynamic phenomena. One mechanism for generating these unstable flows is the Rayleigh-Taylor (RT) instability \cite{Taylor22031950,Rayleigh1883}. 

The RT instability occurs whenever a fluid accelerates another fluid of higher density. The RT growth of interfacial perturbations is driven by the vorticity that is generated by opposed density and pressure gradients. In the (classical) linear regime, each perturbation mode develops independently and grows exponentially \cite{Taylor22031950,Rayleigh1883}. For ablatively driven fronts, this growth is reduced or even stabilized \cite{takabe_PoF_1985, betti_POP_1998}. When the mode amplitude becomes comparable to its wavelength, the modulations develop into bubbles and spikes, where less-dense material rises through heavier material and heavier material falls through less-dense material \cite{McCrory__PRL_1981}. In the nonlinear regime, bubbles merge and evolve self-similarly \cite{Alon1994,Sadot2005} prior to more complex and less ordered behavior that may develop at later times.

When a high-power laser irradiates a solid target, the surface is rapidly heated and the low-density ablated plasma pushes on the higher-density target, forming an ablation front that is RT unstable \cite{lindl1995}. Ablative RT instability growth has been studied extensively because of its relevance to ignition target designs in inertial confinement fusion \cite{knauer:338} and material strength studies at high energy densities \cite{Park_PRL_2010}. In these conditions, uncertainties exist in the magnetohydrodynamic response of high energy density plasma to the generated vorticity \cite{Srinivasan_POP_2102}. Uncertainties include the rate of RT-generated magnetic field growth \cite{Mima1978, Srinivasan_PRL_2012} and the effect of these fields on energy transport \cite{Evans1986, Haines1997}. Previous work showed that magnetic fields could be generated at a laser-driven ablation front by the linear \cite{Manuel2012} and nonlinear \cite{Gao2012} RT instability, but the data were unable to provide detailed maps of the magnetic field spatial distribution, preventing a quantitative comparison with nonlinear RT model predictions. 

This Letter reports measurements of the magnetic field spatial distribution that is generated at a laser-driven ablation front by the nonlinear RT growth of laser-seeded three-dimensional broadband modulations. Planar-target foils were irradiated with $\sim$4-kJ, 2.5-ns laser pulses focused to $\sim$10$^{14}$ W$/$cm$^{2}$ on the OMEGA EP Laser System \cite{Waxer2005}. An ultrafast proton beam measured the growth of cellular magnetic field structures that were generated by the unstable targets. These experiments show, for the first time, self-similar behavior in the RT-generated magnetic fields that grow during target acceleration. The data are consistent with a bubble competition and merger model \cite{Alon1994,oron1998,oron2001}, linking the cellular magnetic field structures with RT bubble and spike growth. The model predicts the time evolution of the number and size of the bubbles, based on the assumption that the bubble-size distribution is self-similar and that the scaled average merging rate is invariant in time. Magnetic field cell-merging rates were measured and are consistent with the RT bubble-merger rates that were determined by earlier x-ray measurements \cite{Sadot2005}. The present work represents a completely independent confirmation of the conclusions drawn in Ref. [15], and is important because of how critical the understanding of RT instability is to inertial confinement fusion \cite{Atzeni2004,lindl1995}. 

\begin{figure}[t]
\includegraphics[width=8cm]{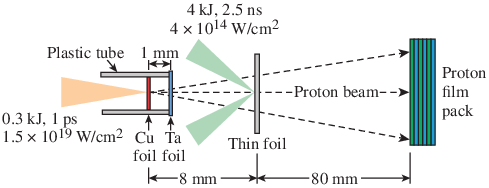}
\caption{\label{fig1} (color online). Experimental setup.}
\end{figure}

Figure 1 shows a schematic of the experimental setup. Two long-pulse beams were used to drive the main target interaction. Each long-pulse beam had a 2.5-ns square temporal profile at a wavelength of 351 nm and delivered 2 kJ of laser energy at a 23$^{\circ}$ angle of incidence to the target normal. The laser beams were focused to 850-$\mu$m-diam focal spots and included distributed phase plates \cite{DPP_LLEReview}. The overlapped laser intensity was 4 $\times$ 10$^{14}$ W$/$cm$^{2}$. The main targets were 15-$\mu$m-thick CH, 5 $\times$ 5 mm$^{2}$ in area.

An ultrafast laser-driven proton source probed the magnetic field spatial distribution inside the plasma and tracked its evolution (see Fig. 1). The proton beam was generated by a 0.3-kJ, 1-ps laser pulse interacting with a 20-$\mu$m-thick Cu foil at focused intensities of 1 $\times$ 10$^{19}$ W$/$cm$^{2}$. This short-pulse beam had a wavelength of 1.053 $\mu$m and was focused with a 1-m-focal-length, \textit{f}/2 off-axis parabolic mirror at normal incidence to the Cu foil. The proton beam was generated by target normal sheath acceleration \cite{wilks2001}. A 5-$\mu$m-thick Ta foil protected the Cu foil from the coronal plasma and x-ray preheat that was generated by the main target interaction. A plastic tube was used to hold these foils, with the Cu foil mounted inside the tube and the Ta foil attached to the side facing the main target. The distance between the Cu foil and the Ta foil was 1 mm. The tube extended an additional 2 mm beyond the Cu foil for further protection against the ablated plasma flow from the main target \cite{zylstra:013511}. The outer and inner diameters of the tube were 1.6 mm and 1.4 mm, respectively. An energetic proton beam with a smooth spatial profile was measured when there was no main target interaction. 

The energetic protons probed the main target interaction and were recorded by a stack of radiochromic film interleaved with aluminum filters. Protons with different energies deposited energy inside different film layers corresponding to their energy-dependent Bragg peak. The probe time measured by each film layer was the sum of the timing difference between the long-pulse and the short-pulse beams and the proton transit time to the main target interaction.  As a result, each film layer measured information about the main target interaction at different times. The image magnification was $M=(L + l)/l$, where $l$ was the distance between the proton-source foil and the main target, and $L$ was the distance between the main target and the radiochromic film detector. In these experiments, $M$ was 11-13 depending on the film layer. This technique provided $\sim$5- to 10-$\mu$m spatial resolution and a temporal resolution of a few picoseconds.

\begin{figure}[t]
\includegraphics[width=7cm]{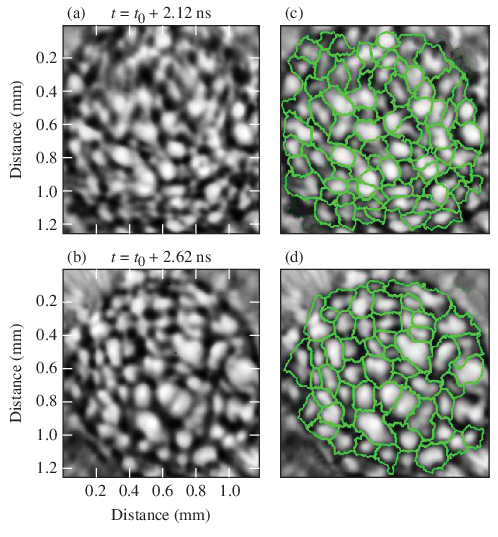}
\caption{\label{fig1} (color online). Proton radiographs of 15-$\mu$m-thick CH foils taken with 25-MeV protons at (a) $t = t_{0}+2.12$ ns and (b) $t = t_{0}+ 2.62$ ns. The magnetic field cells identified with the watershed algorithm are overlaid in (c) and (d).}
\end{figure}

Figure 2 shows proton radiographs of the cellular magnetic field structures that grew as the target was accelerated. Figures. 2(a) and 2(b) show data from two different shots for 15-$\mu$m-thick CH at times $t = t_{0} + 2.12$ ns and $t = t_{0} + 2.62$ ns, where $t_{0}$ is the arrival time of the long-pulse beams at the main target surface. The data were obtained with 25-MeV protons. The window size is 1256 $\mu$m $\times$ 1184 $\mu$m. Darker regions in the images correspond to a higher proton flux, revealing proton beam deflections at the driven target. The data show cellular-field structures that grow during the target acceleration. These sharp cellular structures were previously interpreted as caustics \cite{KuglandRSI} and were created by magnetic fields in the unstable plasma. 

The first experiments to measure these fields were reported in Refs. [23, 24]. The laser and target-interaction conditions reported here were the same as those reported in Ref. [24]. The target modulations were seeded by laser nonuniformities and amplified by RT instability growth during the target acceleration. Large density and temperature gradients formed in the unstable targets and spontaneously generated magnetic fields through the $\nabla n_{e}\times\nabla T_{e}$ mechanism \cite{Mima1978}. This interpretation was previously supported by numerical modeling with the two-dimensional resistive magnetohydrodynamic code DRACO \cite{Keller1999,igumenshchev2009}. 

The data reported here show the first detailed maps of these magnetic fields by probing the targets in a face-on geometry. In these experiments, collisional scattering is negligible and the magnetic fields play a dominant role in deflecting the proton probe beam \cite{Gao2012}. Magnetic fields generated by the RT instability create the measured caustics, or cellular structures. A detailed description for caustic formation in this geometry will be presented in a future publication. For magnetic fields above 0.5 MG and proton energies above 9 MeV, the measured size of the cellular structures at a given probing time is insensitive to the magnitude of the magnetic field. The data reported here show the structural evolution of these fields.

\begin{figure}[t]
\includegraphics[width=6cm]{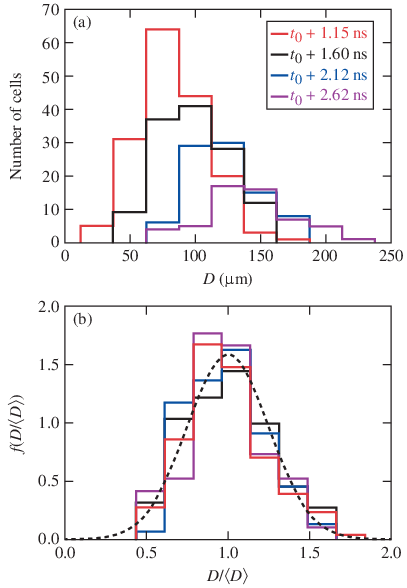}
\caption{\label{fig1} (color online). (a) Magnetic field cell-size distribution for laser-driven CH targets at four different times (see text for details). (b) Normalized magnetic field cell-size distribution as a function of the normalized cell size for each of the cases shown in (a). The data are fitted with a Gaussian distribution function (dashed).}
\end{figure}

The number and size of magnetic field cells that were generated at each probing time were found by analyzing the experimental data with the watershed algorithm \cite{vincent1991watersheds}. Cellular structures near the edge of the interaction window were excluded from the analysis. The cellular structures that were identified with the watershed algorithm are shown in Figs. 2(c) and 2(d), overlaid on the original images. Based on this analysis, the total number of cells was determined and the cell sizes were calculated as $D=(4S/ \pi)^{1/2}$, where $S$ was the cell area, making it possible to construct the cell-size distribution for each of the images. 

Figure 3(a) shows the magnetic field cell-size distribution for four different probing times that were measured on separate shots. In time, the total number of magnetic field cells decreased and the average cell size shifted to longer wavelengths. The cell-size distributions were fitted with normal distributions, from which the average cell sizes $\langle D\rangle$ were determined. At time $t = t_{0}+1.15$ ns, the average cell size was 83 $\mu$m, increasing to 99 $\mu$m, 121 $\mu$m and 143 $\mu$m at times $t = t_{0}+1.60$ ns, $t = t_{0}+2.12$ ns, and $t = t_{0}+2.62$ ns, respectively. These results agree with direct calculations of $\langle D\rangle$ using the cell sizes from the watershed segmentation. Figure 3(b) shows the normalized cell-size distributions as a function of the normalized cell size $D/\langle D\rangle$. The dashed line represents a fit to the experimental data using the Gaussian distribution function
\begin{equation}
 f(D/\langle D\rangle)=\frac{1}{\sqrt{2\pi}\psi}\exp\!\left[-\frac{(D/\langle D\rangle\!-\!1)^2}{2\psi^2}\right]
\end{equation}

\noindent where $\psi=0.25 \pm 0.02$. This analysis shows that the normalized magnetic field cell-size distributions are time invariant and evolve self-similarly \cite{Sadot2005}. 

The RT bubble and spike evolution is revealed by the magnetic fields that were generated by nonuniform electron pressure gradients. Previous DRACO simulations \cite{Gao2012} confirmed that for these plasmas the magnetic pressure is dynamically insignificant. As a result, the spatial distribution and growth of the cellular magnetic field structures are expected to correlate with RT bubble competition and merger model predictions. 

To confirm that the evolution of the cellular magnetic field structures followed the flow-driven system dynamics the data were analyzed using the RT bubble competition and merger model predictions that were outlined in Refs. [14, 15, 26, 27]. The model describes RT instability growth in the nonlinear regime, and predicts that the total number of RT bubbles $N$ decreases in time as
\begin{equation}
\frac{dN}{dt}=-\sqrt{{g}\over{\langle \lambda \rangle (t)}}\varpi N(t)
\end{equation} 

\noindent where $g$ is the target acceleration, $\langle \lambda \rangle (t)$ is the time-dependent average bubble size and $\varpi$ is the dimensionless scaled average bubble-merging rate. The average bubble size growth rate is given by
\begin{equation}
\frac{d\big<\lambda\big>}{dt}=\frac{1}{2}\sqrt{{g}\over{\big<\lambda\big>(t)}}\varpi \big<\lambda\big>(t)
\end{equation}

\noindent The solutions to these equations are 
\begin{eqnarray}
N(t)&=&\phi\!\left[\varpi\sqrt{g}t+2C\right]^{-4},\\[0.25cm]
\langle \lambda\rangle(t)&=&\frac{1}{16}\varpi^2gt^2+\frac{1}{4}C\varpi\sqrt{g}t+\frac{1}{4}C^2,
\end{eqnarray}

\noindent where $\phi =N_0[2C]^4$, $C=2\sqrt{\langle \lambda\rangle_0}$, $N_0$ is the initial number of bubbles, and $\langle \lambda\rangle_0$ is the initial average bubble size. The nonlinearity produces bubble and spike formation, with the interaction between neighboring bubbles governing the modulation evolution. Bubble competition leads to smaller bubbles disappearing and larger bubbles dominating the system dynamics. In time, the average bubble size shifts to longer wavelengths and the bubble size distribution evolves self-similarly. 

\begin{figure}[t]
\includegraphics[width=6cm]{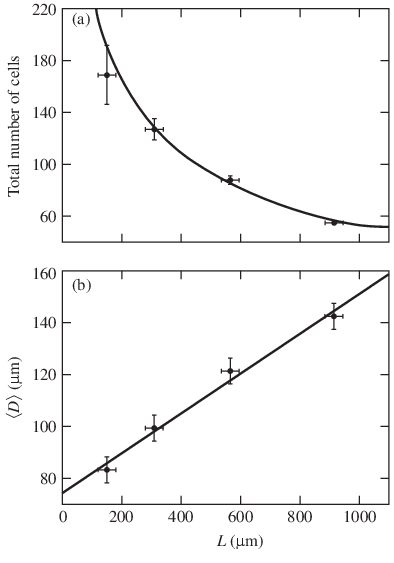}
\caption{\label{fig1} (a) Measured total number of magnetic field cells and (b) average magnetic field cell size as a function of the distance traveled by the target. Solid lines are fits to the experimental data based on bubble competition and merger model predictions \cite{Sadot2005}.}
\end{figure}

The experimental data confirm the scalings that are predicted by the RT bubble competition and merger model, enabling calibration of the scaled magnetic field cell-merging rate. Figures 4(a) and 4(b) show the measured total number of magnetic field cells $N_{c}$ and the average magnetic field cell size $\langle D\rangle$ as a function of the distance traveled by the target. This distance $L=\frac{1}{2}gt^2$, where $g$ is the target acceleration and $t$ is time, determines the amount of RT growth. This distance was measured with side-on proton radiography using the experimental setup described in Ref. [24]. The data were fit to the RT model predictions and compare well with the scaling $N_{c}(t) \propto [\varpi_{c}\sqrt{g}t+2C]^{-4}$ and $\langle D\rangle \propto \varpi_{c}^2gt^2$, where $\varpi_{c}$ is the scaled magnetic field cell-merging rate. The evolution of the average magnetic field cell size compares well with the self-similarity of RT growth, where $\langle \lambda \rangle$ grows proportionally to $gt^2$. Based on this analysis, it was found that $N_{c,0}=556\pm 50$, $\langle D\rangle_0 = 75\pm 5$ $\mu$m, and $\varpi_{c}=0.79\pm 0.06$. Within the experimental error, this merging rate agrees with the value that was measured in previous nonlinear RT instability growth studies with laser-driven plastic targets using x-ray radiography ($\varpi = 0.83 \pm 0.10$) \cite{Sadot2005}.






In summary, detailed maps of the magnetic field cells that are generated by the ablative nonlinear RT instability were measured in laser-accelerated planar targets. The magnetic field evolution shows self-similar behavior, and the observations were quantitatively consistent with a bubble competition and merger model, linking the generated magnetic fields with the RT bubble and spike growth. These observations are a first step towards understanding the evolution of large-scale coherent magnetic field structures that can, under certain circumstances, spontaneously emerge and persist in strongly driven flows at high energy densities. A compelling aspect of this work is the creation of globally coherent magnetic field structures seeded and forced by hydrodynamic instability growth. This could benefit the understanding of magnetic-seed-field generation in high energy density plasmas and the flow-driven processes that induce global magnetic structure prior to their turbulent amplification. In strongly driven plasmas, such information is difficult to obtain by any other technique.

This work was supported by the U.S. Department of Energy Office of Inertial Confinement Fusion under Cooperative Agreement No. DE-FC52-08NA28302, the University of Rochester, and the New York State Energy Research and Development Authority. The support of DOE does not constitute an endorsement by DOE of the views expressed in this article.

\bibliographystyle{unsrt}
\bibliography{1MAIN}

\end{document}